\documentstyle[twocolumn,aps,epsf]{revtex}
\begin{document}
\title{Effect of Disorder on Charge Density Wave and 
Superconducting Order in the Half-Filled Attractive Hubbard Model}
\author{C. Huscroft and R.T. Scalettar}
\address{
Physics Department,
University of California,
Davis, CA 95616}
\address{\mbox{ } {\rm (\today )} }
\address{\mbox{ }}
\address{\parbox{14cm}{\rm \mbox{ }\mbox{ }
The half-filled attractive Hubbard model
exhibits simultaneous charge density wave and superconducting order
in its ground state.  In this paper we explore the effect of
disorder in the site energies on this degeneracy.  We find that 
superconducting order survives randomness out to a critical
amount of disorder, but charge ordering is immediately reduced.
We explore the validity of a strong coupling picture which maps the
system onto a Heisenberg model in a random magnetic field.
}}
\address{\mbox{ }}
\address{\parbox{14cm}{\rm
\noindent PACS numbers: 71.55.Jv,74.20.-z,71.45.Lr,71.27.+a,75.10.Nr}}
\address{\mbox{ }}
\newpage
\maketitle

\section{Introduction}

The repulsive Hubbard model has long served as a simple
Hamiltonian to describe itinerant magnetism. 
Similarly, the attractive Hubbard model has
been used to explore qualitative features of the
superconducting phase transition.\cite{RANDERIA1}
This model does not provide a microscopic 
model of the origin of pairing.  Rather it is assumed that some 
other degrees of freedom, for example an electron-phonon coupling,
have already provided the necessary attraction.
Recent quantum simulation 
studies\cite{SCALETTAR1,FYE1,IMADA1,ASSAAD1,ASSAAD2}
have explored a number of features of the attractive Hubbard
Hamiltonian, including a determination of the superconducting
transition temperature and its dependence on electron density,\cite{MOREO1}
the detailed spatial structure of the pairing correlations,\cite{MOREO2}
the coexistence of a Bose-like spin gap with an otherwise degenerate Fermi
gas of electrons,\cite{TRIVEDI1} and deviations from Fermi-liquid
behavior.\cite{TRIVEDI2}
Much analytic work has also been done, as reviewed in \cite{RANDERIA1}.

While the interplay of superconductivity
and disorder has of course been extensively studied 
theoretically,\cite{REVIEWS}  
considerably less is known numerically.
The competition between the dephasing effect of impurity
scattering and the dramatic manifestation of phase
coherence in the zero resistance state gives rise to a set of
challenging qualitative questions.  A quantitative
understanding is also essential in order to 
model experiments like those which address the question
of the possibility of a universal resistance
in disordered superconducting films.\cite{EXPERIMENTS}

Much recent theoretical\cite{FISHER1,FISHER2}
and numerical\cite{CHA1,SORENSEN1,RUNGE1,SCALETTAR2,TRIVEDI4}
work on these issues has been
done within the context of the ``boson-Hubbard'' model; that is,
under the assumption that preformed Cooper pairs exist even in the
non-superconducting state and the transition is driven by the loss
of phase coherence, rather than the destruction of the
magnitude of the superconducting gap.
This bosonic model should be the limit of the attractive Hubbard 
Hamiltonian as the on-site interaction $U$ becomes large.  Despite the
greater computational simplicity of the boson models, it is the fermion system
which is of fundamental interest.

In this paper we will explore the effect of random site energies in the
attractive Hubbard Hamiltonian:
\begin{eqnarray}
H=-t\sum_{\langle {\bf i},{\bf j} \rangle \sigma} 
(c_{{\bf i}\sigma}^{\dagger}c_{{\bf j}\sigma} + 
c_{{\bf j}\sigma}^{\dagger}c_{{\bf i}\sigma})\nonumber\\
   - |U| \sum_{{\bf i}} (n_{{\bf i}\uparrow}-\frac12)
(n_{{\bf i}\downarrow}-\frac12)\nonumber\\
   + \sum_{{\bf i}} (\epsilon_{{\bf i}}-\mu) 
(n_{{\bf i}\uparrow}+n_{{\bf i}\downarrow}).
\label {eq:eq1}
\end{eqnarray}
Here $c_{{\bf i}\sigma}$ ($c_{{\bf i}\sigma}^{\dagger}$) are operators which
destroy (create) electrons of spin $\sigma$ on site ${\bf i}$,
so the first term in $H$ describes the hopping of electrons between
nearest neighbor sites on our 2d square lattice.
$|U|$ is the on-site attraction, while $\mu$ and $\epsilon_{{\bf i}}$
are the chemical potential and random site energies, respectively.
The 
$\epsilon_{{\bf i}}$ are chosen uniformly on $[-V,+V]$.  In this paper,
we will work exclusively at half-filling, $\langle n_{{\bf i}\uparrow}
+n_{{\bf i}\downarrow} \rangle=1$.
A related study of the current-current correlations
and the behavior of the resistivity as a function of
disorder strength and temperature away from half-filling
is contained in \cite{TRIVEDI3}.

In the absence of disorder, $V=0$, considerable insight can be gained
by considering the effect of a particle-hole transformation on
the down electron operators,
$c_{{\bf i}\uparrow} \leftrightarrow c_{{\bf i}\uparrow}, \hskip0.1in
c_{{\bf i}\downarrow} \leftrightarrow c_{{\bf i}\downarrow}^{\dagger} 
(-1)^{i_{x}+i_{y}}.$
The phase factor $(-1)^{i_{x}+i_{y}}$ changes sign as one goes between the two
sublattices of our (bipartite) square lattice.
Under this transformation the kinetic energy is invariant,
while the interaction changes sign, $|U| \leftrightarrow -|U|$.
The chemical potential now couples to the $z$ component
of spin on each site, instead of to charge.  In the absence of
a chemical potential term, we have an exact mapping between the
attractive and repulsive Hubbard Hamiltonians.  
Pair correlations in the attractive model map onto
spin correlations in the $xy$ plane of the repulsive model, while
cdw correlations are associated with spin correlations in the $z$ direction.
Since the long range spin order in the ground state
of the 2D repulsive Hubbard model is rotationally invariant,
we immediately conclude that pairing and cdw correlations coexist in
the ground state of the half-filled attractive model.
This mapping can also be used to discuss the effect of doping,
as will be described in section 5.

In this manuscript we consider non-zero disorder, $V \neq 0$.
Our basic conclusions are:
(i)  The addition of site disorder breaks the degeneracy
between superconducting and charge-density wave states.
Equal time density-density correlations exhibit a 
rapid suppression of their staggered pattern, while
pair-pair correlations remain robust.
(ii)  Our data is consistent with an immediate destruction of
long range charge order, while superconducting correlations 
appear to persist out to a finite $V_{c} \approx 1.5$.
(iii)  The behavior of superconducting correlations
at $|U|=4$ are in approximate agreement with that of the
appropriate magnetic correlations in a strong coupling model.

An outline of the remainder of this paper is as follows:
In section 2 we give a brief description of the
simulation and of the observables used to characterize the 
ground state correlations.  In section 3 we present
results for local pair and charge correlations.
Section 4 provides a finite size scaling
analysis of this data to determine the existence of long range order.
We also discuss in some detail
the distribution of measurements for different
disorder realizations.
Section 5 discusses results for the strong coupling version of this
model, the antiferromagnetic Heisenberg model in
a random magnetic field.  A summary is presented in
section 6.

\section{Brief Description of the Simulation}

We will use the ``determinant'' monte carlo algorithm\cite{SUGAR1}
for our numerical work.  In this approach the partition
function is written down as a path integral by discretizing
the inverse temperature $\beta$ and using the Trotter
approximation\cite{TROTTER1,SUZUKI1}
to break up the exponential of the kinetic
and potential pieces of the Hamiltonian.
The interaction term is decoupled with a discrete Hubbard-Stratonovich
transformation\cite{HIRSCH1}.  The resulting trace over
the fermion operators is over the exponential of quadratic
forms, and so can be done analytically.
The result is an expression for the partition function
which is the sum over all values of the discrete Hubbard-Stratonovich
field of a summand which is the product of two determinants,
one each arising from the spin up and the spin down degrees of freedom.
Because both the random site energies and the auxiliary field
couple to the charge, the two determinants are identical
and hence their product is positive.  There is no fermion
sign problem in these simulations, even away from
half filling.

The matrices $M_{\sigma}$ whose determinants give the weight of a particular
auxiliary field configuration are simply related to the equal time fermion 
Green function:  $G_{i j}\equiv\langle c_i c^\dagger_j \rangle =M^{-1}_{i j}$.  Observables are measured by 
expressing them (using Wick's theorem) in terms of the
appropriate sums and products of $G$.
Of particular interest to us here are the equal time
charge and pair correlations,
\begin{eqnarray}
c({\bf j} - {\bf l}) &=& \langle \,\, (n_{\uparrow {\bf l}} 
+ n_{\downarrow {\bf l} }-1)
(n_{\uparrow{\bf j}} + n_{\downarrow {\bf j} }-1) \,\, \rangle,
\nonumber\\
p_s ({\bf j} - {\bf l}) &=& \langle \,\, \Delta_{{\bf l}}^{ \,}
\Delta_{{\bf j}}^{\dagger} \,\, \rangle,
\nonumber\\
\Delta_{{\bf j}}^{\dagger}  &=& c_{\uparrow {\bf j}}^{\dagger}
c_{\downarrow {\bf j}}^{\dagger},
\label {eq:eq2}
\end{eqnarray}
and their associated structure factors,
\begin{eqnarray}
S_{{\rm cdw}} &=& {1 \over N} \sum_{{\bf j},{\bf l}} c({\bf j} - {\bf l}) 
(-1)^{|{\bf j} - {\bf l}|},
\nonumber\\
S_{{\rm pair}} &=& {1 \over N} \sum_{{\bf j},{\bf l}} p_s ({\bf j} - {\bf l}). 
\label {eq:eq3}
\end{eqnarray}

\begin{figure}
\begin{center}
\leavevmode
\hbox{%
\epsfxsize=3.0in
\epsffile{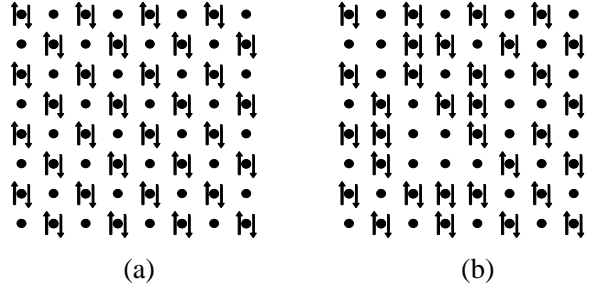}}
\end{center}
\caption{Two possible real space configurations of the electrons
on the lattice are shown.  The pair structure factor attains
its largest value $N/4$ for both,
while the cdw structure factor is
large only for (a).}
\label{FIG1}
\end{figure}

It is useful to look at a few strong coupling snapshots of possible
real space electron distributions to gain a preliminary insight into
these correlation functions and
the effect of site disorder.
Consider the strong coupling limit when all electrons in the system
are paired.  A typical low energy state consists then of a lattice
with each site either empty or doubly occupied.  A configuration in
which the doubly occupied sites alternate with empty sites 
(Fig.~1a) has a lower
energy than one in which doubly occupied or empty sites are
adjacent by an amount $\Delta E \propto -t^{2} / U $ to second order in
the hopping $t$.  (In the language of the repulsive model onto which
the attractive model maps via the particle-hole transformation
discussed above, this energy lowering stabilizes antiferromagnetism
over ferromagnetism at half-filling, and is proportional to the exchange
constant $J$.)
The charge density and pair structure factors defined in Eq.~3 take on
their maximal values ($N/4$ for $S_{{\rm pair}}$ and $N$ for 
$S_{{\rm cdw}}$).  

Now consider the effect of site disorder.  It will not lead to the 
breaking of the pairs, but, when the site energies exceed
$\propto t^{2} / U$, it will change the sites on which
the pairs prefer to reside (Fig.~1b).  Note that the pair structure
factor is still large for such a disordered configuration of pairs, since
contributions to it depend only on finding doubly occupied sites
and empty sites somewhere in the lattice to which to hop.
However, the phases in the charge structure factor make it
extremely sensitive to the destruction of the original staggered
pattern.  This rough argument suggests that
pair order will be more robust to randomness in the
site energy than will charge ordering.
Of course, on general grounds we also expect a term in the
Hamiltonian which couples directly to the charge to have
the greatest effect on the associated charge correlations.
Indeed, as discussed by Anderson \cite{ANDERSON1}
non-magnetic impuritites are not expected to destroy superconductivity,
since one can still pair appropriate eigenstates of the single particle
Hamiltonian which includes the randomness.  Even when the disorder is
large enough to localize these eigenstates, it has been suggested
that superconductivity survives.\cite{MA&LEE}

This overview captures the essence of how disorder
affects our system.  In the next section, we
will make this qualitative picture more precise.

\section{Local Correlations} 

We begin by showing some results for the disorder dependence of
local quantities.  The kinetic energy
$\langle k \rangle=-{t \over N} \sum_{\langle {\bf i},{\bf j} \rangle \sigma} 
(c_{{\bf i}\sigma}^{\dagger}c_{{\bf j}\sigma} + 
c_{{\bf j}\sigma}^{\dagger}c_{{\bf i}\sigma})$
is shown
in Fig.~2.  As we shall see below, superconductivity
vanishes around $V_{c} \approx 1.5$.  The kinetic energy 
shows no special signal at this transition.  Of course
a measure of local electron hopping like
$\langle k \rangle$ does not have to vanish at an insulating
phase transition.\cite{BATROUNI1}

\begin{figure}
\begin{center}
\leavevmode
\hbox{%
\epsfxsize=3.25in
\epsfysize=2.75in
\epsffile{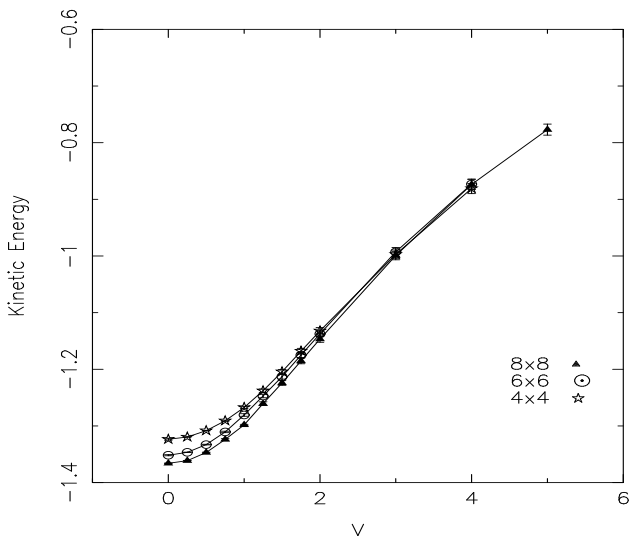}}
\end{center}
\caption{The kinetic energy as a function of disorder strength.}
\label{FIG2}
\end{figure}

\begin{figure}
\begin{center}
\leavevmode
\hbox{%
\epsfxsize=3.25in
\epsfysize=2.75in
\epsffile{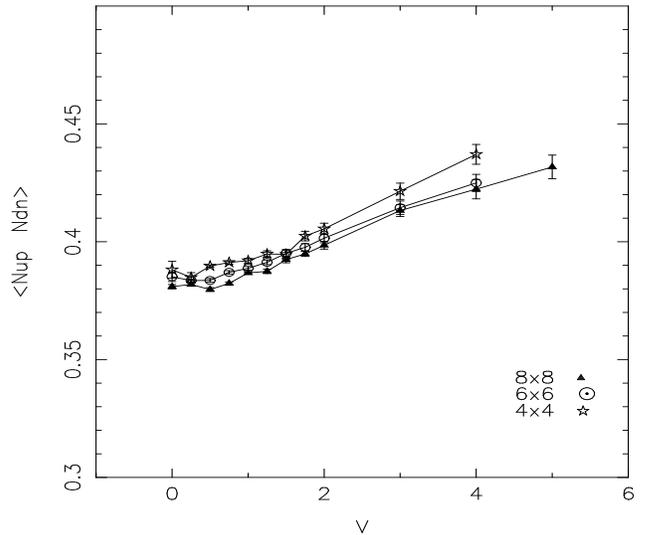}}
\end{center}
\caption{The double occupancy rate as a function of disorder strength.}
\label{FIG3}
\end{figure}

\begin{figure}
\begin{center}
\leavevmode
\hbox{%
\epsfxsize=3.25in
\epsfysize=2.75in
\epsffile{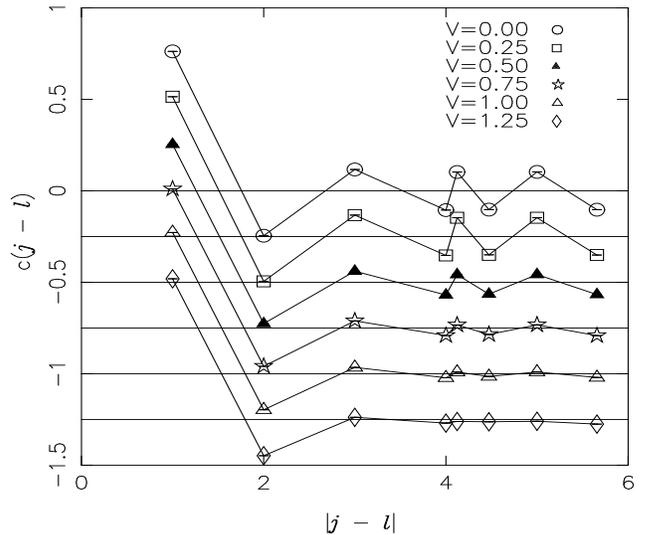}}
\end{center}
\caption{Charge correlations as a function of site separation
for different disorder strengths.  Successive disorder strengths have been offset vertically for clarity.  The horizontal lines indicate the c({\bf j} - {\bf l})=0 axis for each successive disorder strength.  (The numbers labeling the vertical axis correspond to the V=0.00 case.)}
\label{FIG4}
\end{figure}

\begin{figure}
\begin{center}
\leavevmode
\hbox{%
\epsfxsize=3.25in
\epsffile{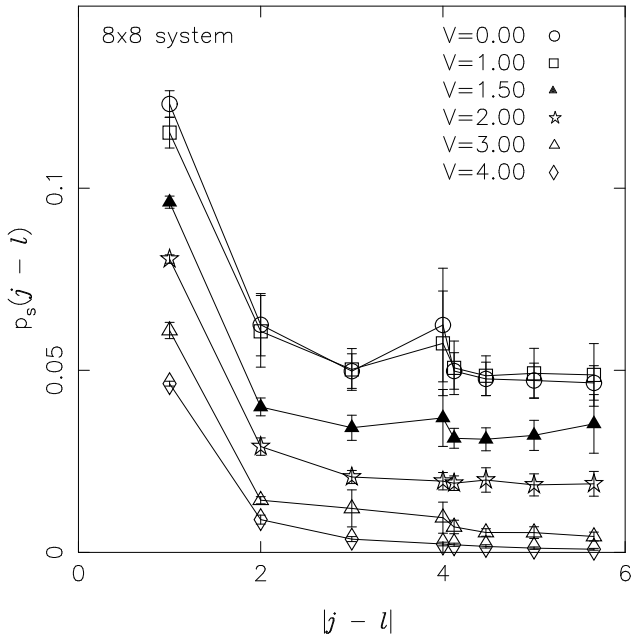}}
\end{center}
\caption{Pair correlations as a function of site separation
for different disorder strengths.}
\label{FIG5}
\end{figure}

\begin{figure}
\begin{center}
\leavevmode
\hbox{%
\epsfxsize=3.25in
\epsffile{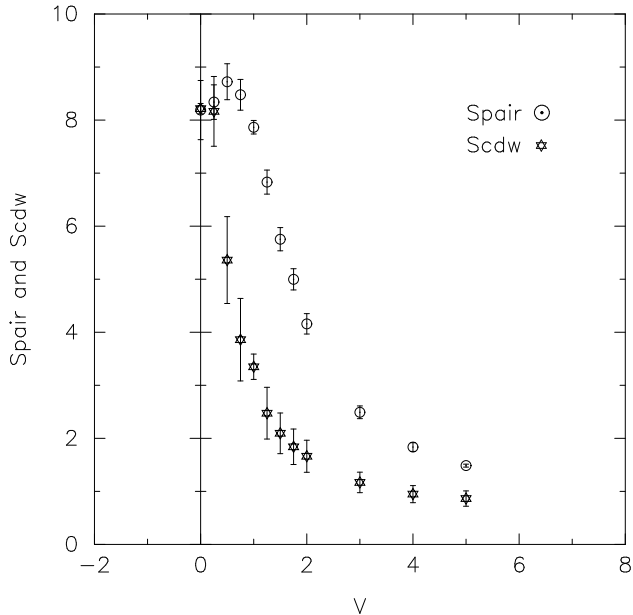}}
\end{center}
\caption{Charge and pair structure factors as a function of disorder
strength.}
\label{FIG6}
\end{figure}

In the repulsive Hubbard model, random site energies have 
a fundamental qualitative effect on the double occupancy rate
$\langle n_{{\bf j} \uparrow} n_{{\bf j} \downarrow} \rangle$
since they compete with the
repulsive interaction and disturb local moment formation.
In the attractive model,
we expect site disorder to have a much less dramatic effect
and indeed that is seen to be the case in Fig.~3.

Longer range charge-charge correlations are dramatically affected
by site disorder.  Fig.~4 shows $c({\bf j} - {\bf l})$ as a function of lattice
separation ${\bf j} - {\bf l}$ for different disorder strengths.  
(Successive disorder strengths have been offset vertically for clarity.)
The lattice size is 8x8, inverse temperature $\beta=10$
and $|U|=4$.  The oscillatory character of the charge
correlations is indicative of cdw ordering.  At $V=0$
these correlations extend over the entire lattice, that is,
the correlation length $\xi$ exceeds the linear lattice dimension.
However as $V$ is turned on the correlations go to zero.

Fig.~5 shows the analogous plot for the pair correlation
function $p_s ({\bf j} - {\bf l})$.  The pair correlations remain unchanged for
weak $V$, then are eventually suppressed for sufficiently 
large randomness.  We see the robustness of the pair correlations
as compared to
charge-charge correlations in this plot, which has a much greater range of
disorder strengths, $V,$ than does Fig. 4.

The Fourier transforms of these real space correlation functions
are shown as a function of disorder in Fig.~6.
The degeneracy between charge and pair correlations is
evident in the absence of randomness, $V=0$.
As was seen in Figs.~4 and 5, nonzero site disorder more
rapidly destroys the charge density wave
than the pair correlations.

\section{The Destruction of Long Range Order} 

To determine whether ground-state long range order exists in our system
we need to do a finite size scaling analysis.  
As has been discussed\cite{HIRSCH2,WHITE1}
within the context of the repulsive Hubbard model,
spin-wave theory\cite{HUSE1} predicts that on a 2D lattice
of size N=LxL the charge structure factor and
correlation function at largest separation should behave as
\begin{eqnarray}
{1 \over N} S_{{\rm cdw}} &=& m^{2}/3 +a/L
\nonumber\\
c(L/2,L/2) &=& m^{2}/3  + b/L
\label {eq:eq4}
\end{eqnarray}
in the ordered phase.
Similar results are valid for the pair correlations.
Here $m$ is the order parameter.
Thus, in the ordered phase a plot of the scaled structure factor
versus $1/L$ should be
a straight line with a nonzero intercept
giving $m^2 / 3$.  We will always choose the inverse temperature
$\beta$ sufficiently large that we are effectively at $T=0$ for our finite
lattices.

\begin{figure}
\begin{center}
\leavevmode
\hbox{%
\epsfxsize=3.25in
\epsfysize=2.75in
\epsffile{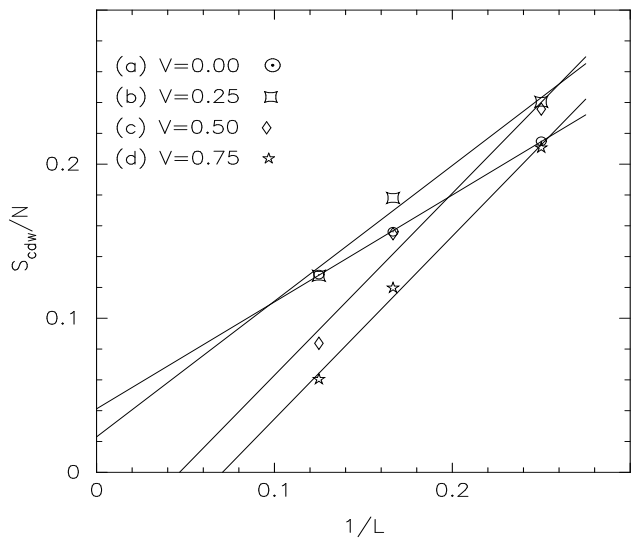}}
\end{center}
\caption{Finite size scaling plots for the charge correlations:
(a) V=0.0; (b) V=0.25; (c) V=0.50; and (d) V=0.75.  The straight lines
are least-squares fits to the data.  Error bars (not shown) on the V=0.25 case are consistent with a zero intercept, while error bars on the V=0.00 case are
not consistent with a zero intercept.  The V=0.00 (clean case) error bars are
much smaller than in the disordered V=0.25 case, since there is no disorder
averaging required in the clean case.}
\label{FIG7}
\end{figure}

\begin{figure}
\begin{center}
\leavevmode
\hbox{%
\epsfxsize=3.25in
\epsfysize=2.75in
\epsffile{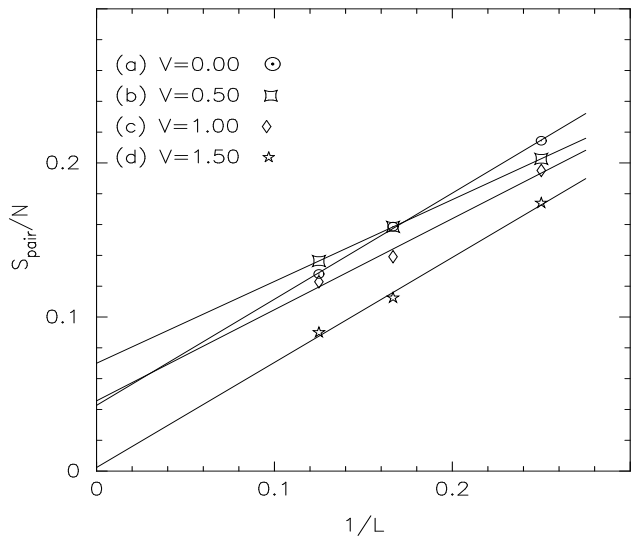}}
\end{center}
\caption{Finite size scaling plots for the pair correlations:
(a) V=0.0; (b) V=0.50; (c) V=1.00; and (d) V=1.50.  The straight lines
are least-squares fits to the data.}
\label{FIG8}
\end{figure}

Fig.~7 shows the result of this analysis for the charge correlations.
The interaction strength is $|U|=4$.
Only in the clean system at $V=0$ is a nonzero order parameter $m$
obtained.  However, as seen in Fig.~8 the pair field order parameter
remains nonzero out to approximately $V=V_{c} \approx 1.5$.\cite{ERRORBARS}

\begin{figure}
\begin{center}
\leavevmode
\hbox{%
\epsfxsize=3.25in
\epsffile{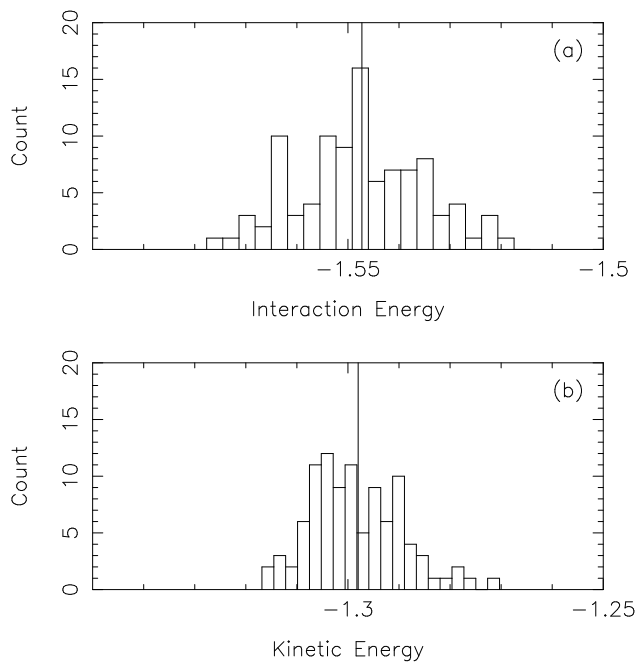}}
\end{center}
\caption{Histogram of values of the (a) interaction and (b) kinetic
energies for different disorder realizations.  Each plot represents data from one hundred disorder realizations.  The average values are shown by the vertical lines.}
\label{FIG9}
\end{figure}

We will conclude this section with a discussion of disorder averaging,
since while the other aspects of our simulation are identical to
those long reported for determinant monte carlo, little is known about
what happens when randomness is included.
In Fig.~9 we show some histograms of the interaction and kinetic
energies for an 8x8 spatial
lattice at inverse temperature $\beta=10$.  We have chosen $|U|=4$ and
$V=1$.  We see that these quantities have a fairly sharp distribution,
that is, the energy is not too sensitive to the detailed disorder 
realization; the widths of the distributions are less than 5 \% of the 
average values.  The error bars associated with realization to realization
fluctuations in these quantities are roughly ten times
the statistical uncertainties in a run consisting of
1000 warm-up sweeps and 5000 measurement sweeps for a single realization.

\begin{figure}
\begin{center}
\leavevmode
\hbox{%
\epsfxsize=3.25in
\epsffile{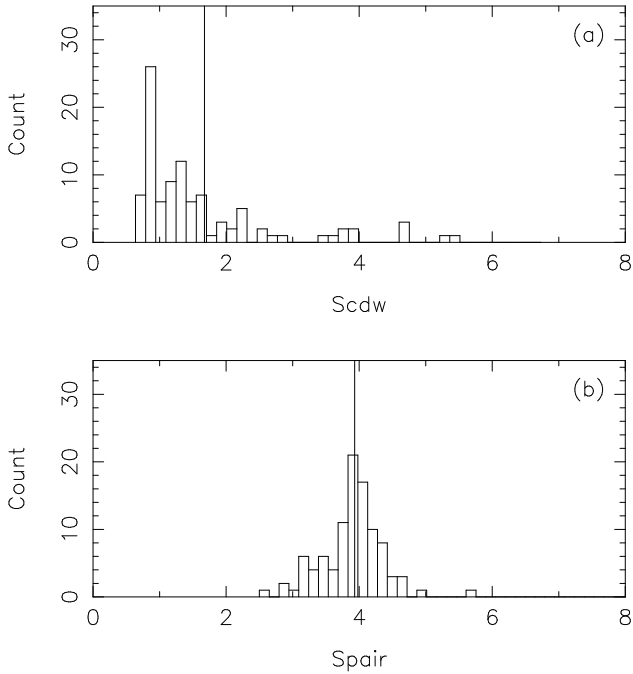}}
\end{center}
\caption{Histogram of values of the (a) charge and (b) pair structure factors
for different disorder realizations.  Each plot represents data from 100 disorder realizations.  The average values are shown by the vertical lines.  The charge structure factor, $S_{cdw},$ has an apparently non-gaussian form.}
\label{FIG10}
\end{figure}

On the other hand, the fluctuations in quantities which measure long range
correlations are, as expected, much larger.  In Fig.~10 we show
histograms of the charge and pair structure factors for an 8x8 spatial
lattice at inverse temperature $\beta=10$, with $|U|=4$ and
$V=1$.  We see that the widths
of the peaks are of the same order of magnitude as the average values of the
respective structure factors.

We note that practical difficulties limit the number of disorder realizations,
and consequently the reduction in 
fluctuations due to disorder averaging, that one can
attain.  A single disorder realization for an 8x8 system at $\beta = 10$
takes over one-thousand cpu minutes on a fast workstation.  This computational
difficulty of these QMC calculations precludes disorder averaging over 
thousands of realizations, as has been done in the spin glass 
literature.  
The non-gaussian nature
of the distributions ({\it e.g.,} Fig.~10a) of course raises 
difficult questions about how to 
do the averaging and how to estimate error bars correctly.  However, if 
one goes ahead and employs the usual methods of getting error bars 
based on an assumption
of a gaussian distribution, then averaging over 20-100 disorder realizations
reduces the statistical errors to about the same level as the 
statistical errors associated with the monte carlo sampling.
This is what we have done in the data reported in 
this paper.

\section{The Random Field Heisenberg Model}

As we have discussed, the attractive ($-U$) Hubbard model can be mapped onto
the repulsive model, which in turn at strong coupling can be mapped
onto a quantum spin 1/2 antiferromagnetic Heisenberg Hamiltonian.
In the absence of disorder, the behavior of the associated 
classical spin models has exhibited considerable analogies to the
original $-U$ Hubbard model. \cite{SCALETTAR1}  Here we desire 
to see if similar
connections can usefully be made between the disordered, $-U$
Hubbard model and the associated classical model -- the random field
Heisenberg model.  However, the problem of the random
field Heisenberg model is an extremely difficult one in it own right.
We emphasize that we are attempting only qualitative contact with the
attractive Hubbard model simulations here.

We begin by reviewing the results in a uniform magnetic field,
since the comparison will be useful in discussing
the case of a random field.  Similar results 
were presented in \cite{SCALETTAR1}.
However, here we present some additional
plots which help more precisely characterize the 
nature of the ordered phase.  In the absence of
a field, the continuous symmetry of the model assures us that
in 2D there can be no true long range order except in the
ground state $T=0$.\cite{MERMIN1}  If a field $h_{z}=\mu$
is applied, the spins tend to lie down in the $xy$ plane, because then
they can tilt upwards in the $z$ direction and take advantage of the
field energy without costing as much exchange energy $J$ as if they
were antiferromagnetically aligned in the $z$ direction.
Thus, the antiferromagnetic Heisenberg model in a uniform magnetic
field is argued to be in the universality class of the
$XY$ model, with a finite temperature Kosterlitz-Thouless phase transition
into a state with power law decay of the correlation functions.
In the language of the attractive model, doping breaks the cdw-pair
degeneracy, and off half-filling one has a finite temperature
phase transition into a purely superconducting state. \cite{MOREO1}

Let us define the antiferromagnetic structure factors in the different
spin directions $\alpha = x,y,z$ as the appropriate
sums of the correlations of the $\alpha$ component
of spin on different sites,
\begin{equation}
S_{{\rm \alpha \alpha}} = {1 \over N} \sum_{{\bf j},{\bf l}} 
\langle s_{\alpha}({\bf j}) 
s_{\alpha}({\bf l}) \rangle
(-1)^{|{\bf j} - {\bf l}|}.
\label {eq:eq5}
\end{equation}

\begin{figure}
\begin{center}
\leavevmode
\hbox{%
\epsfxsize=3.25in
\epsffile{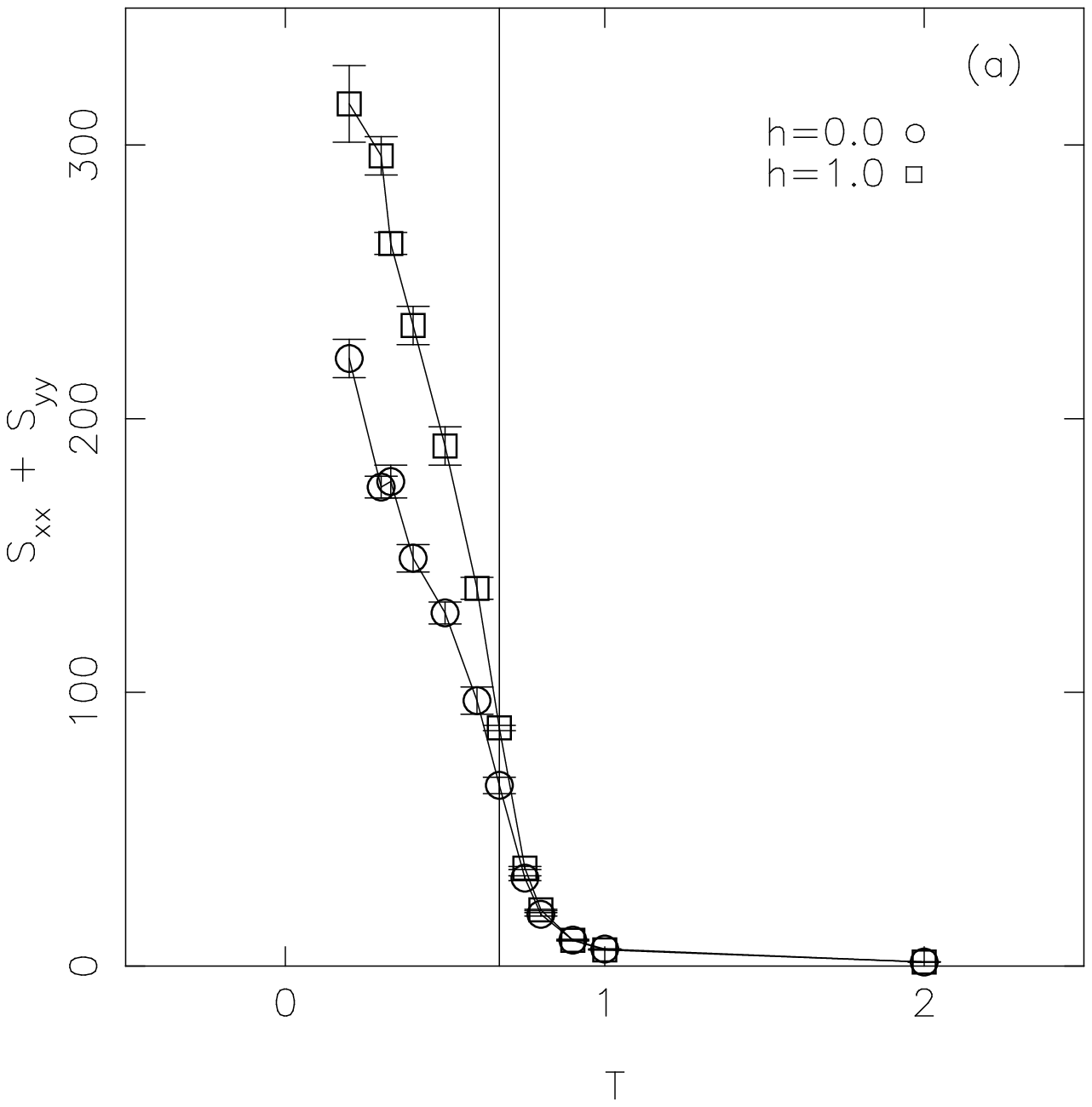}}
\end{center}
\end{figure}

\begin{figure}
\begin{center}
\leavevmode
\hbox{%
\epsfxsize=3.25in
\epsffile{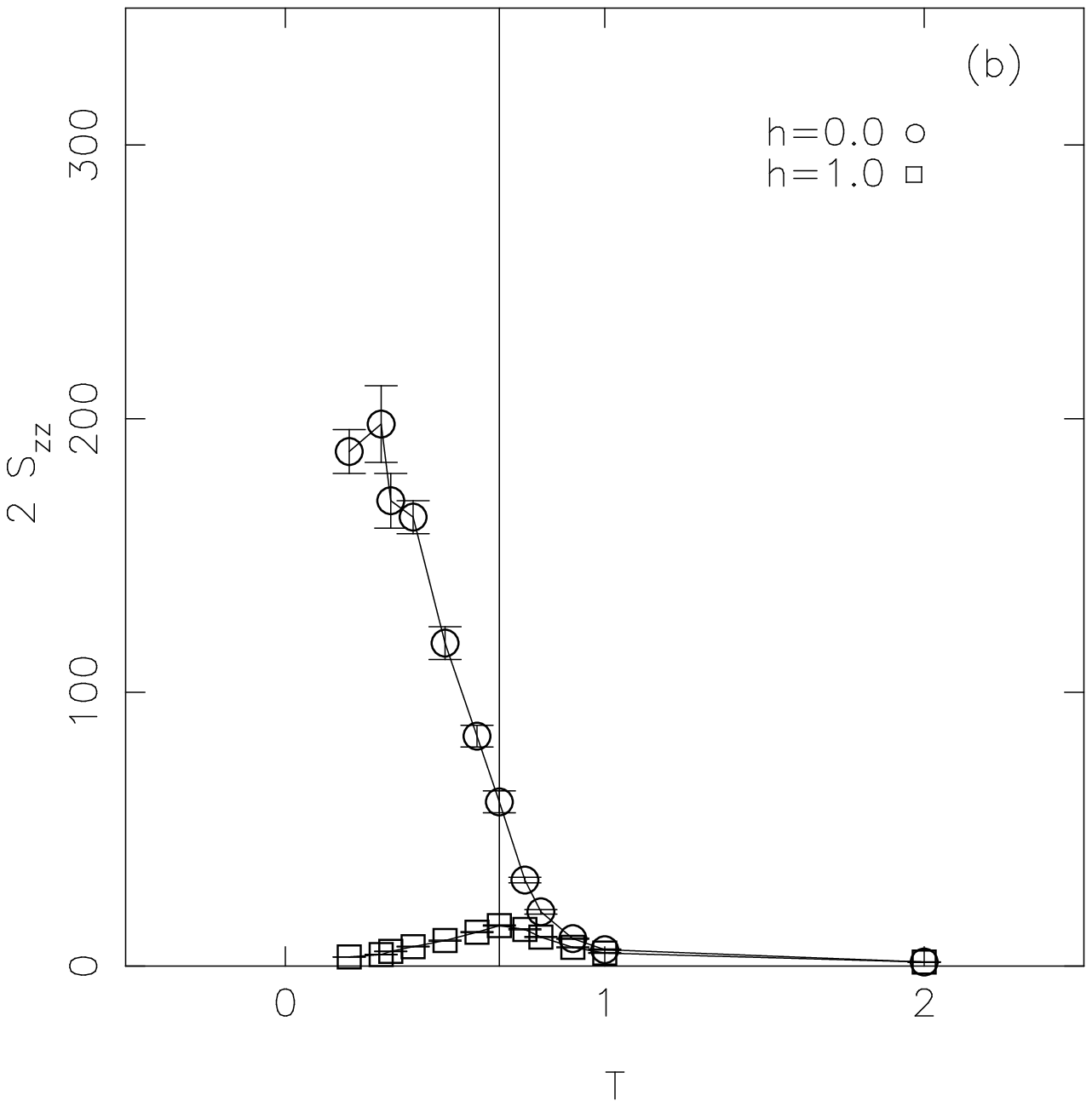}}
\end{center}
\caption{Results for the structure factors as a function of temperature
for a 20x20 lattice.  (a) $S_{xx} + S_{yy};$ (b) $2 S_{zz}.$  The line at
T=0.67 is at the temperature used in the finite size scaling plot (Fig.~12)
 below.}
\label{FIG11}
\end{figure}

\begin{figure}
\begin{center}
\leavevmode
\hbox{%
\epsfxsize=3.25in
\epsfysize=2.75in
\epsffile{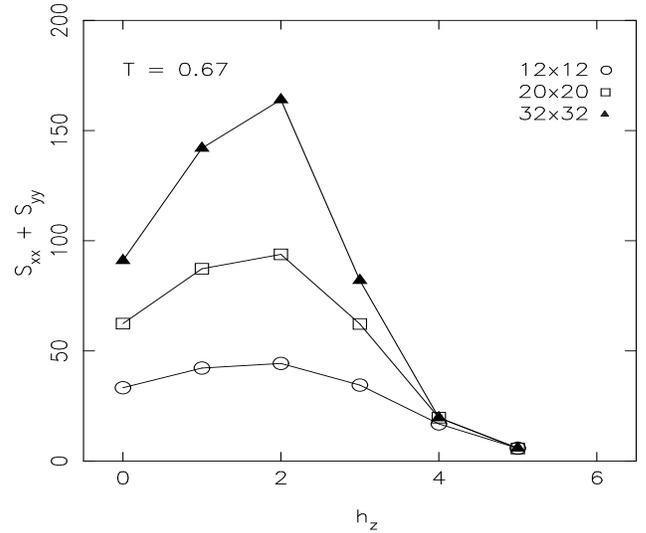}}
\end{center}
\caption{Results for the $xy$ structure factors as a function of uniform
field $h_{z}$ for different lattice sizes.}
\label{FIG12}
\end{figure}

Simulation results for the {\it classical} antiferromagnetic Heisenberg model
in a uniform magnetic field are shown in Figs.~11-13.
The temperature dependence is shown first in Fig.~11 on a fixed lattice size.
To set a possible scale of $T$, note that $T_{{\rm KT}} = 0.725$
for the XY model.\cite{LANDAU}  
This is consistent with the temperature at which the structure factor
swings upward in Fig.~11.
We note that a small uniform field, $h_{z} = 1.0$, enhances  
$S_{xx}$ and $S_{yy}$ substantially, and dramatically reduces $S_{zz}.$
Fig.~12 shows the field dependence of the $xy$ structure factors
at a fixed $T=0.67$ 
for different lattices.\cite{CAVEAT}
When $h_{z}$ is nonzero, there is a significant size dependence of the
structure factor even at nonzero temperature, which
suggests that the presence of a field may indeed make the system
order at finite temperature.

Of course this data is only suggestive.  A careful finite size scaling
analysis would be needed to pin down
whether $T_{c}=0$ or $T_{c}\ne0$.
To illustrate one of the issues involved,
we note that even
if a phase transition occurs only at $T=0$, structure factors will show
significant size dependence up to the lattice size, $L \approx \xi$. 
In the Heisenberg model, the corrrelation length 
$\xi = C_{\xi} \exp(2 \pi J / T) / (1 + 2 \pi J / T)$ 
where $C_{\xi} \approx 0.01.$ \cite{SHENKER1}
Here, with $T = 0.67$, the correlation length $\xi \approx 11$.
The growth in the structure factors $S_{\alpha \alpha}$
with lattice size at $h_{z} = 0$ is consistent with this value of $\xi$,
and hence with $T_{c} = 0$, as should be the case.
The increased growth of $S_{\alpha \alpha}$ with lattice size at nonzero
$h_{z}$ could merely reflect a larger $\xi$ (but $T_{c}$ still zero)
or a finite $T_{c}$.
The structure factors, $S_{\alpha \alpha}$, continue to show the effect 
of this finite $\xi$ throughout the range of lattice sizes used in the
simulations illustrated in Fig.~12.
Hence, the value of $\xi$ in this
model is consistent with 
the results shown in Fig.~12 being at a temperature where correlations
have started to form across the lattice.

\begin{figure}
\begin{center}
\leavevmode
\hbox{%
\epsfxsize=3.25in
\epsfysize=2.75in
\epsffile{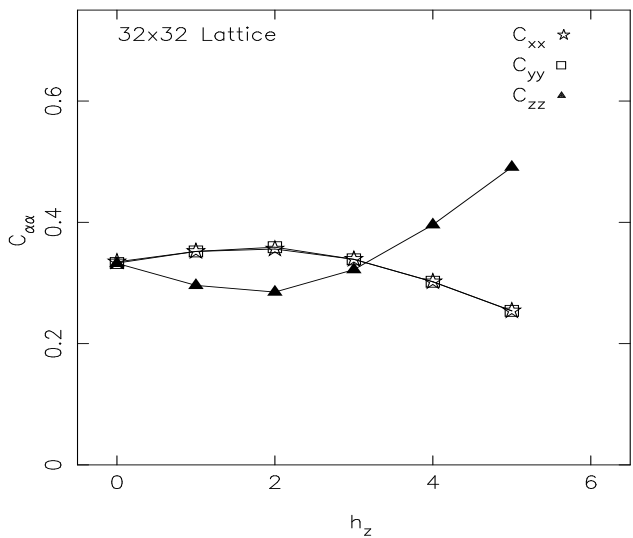}}
\end{center}
\caption{The averages of the squares of different spin components
$C_{\alpha \alpha} = \langle s_{\alpha}^{2} \rangle$ as a function of
uniform field $h_{z}$.  For $h_{z}=0,$ $C_{x x}=C_{y y}=C_{z z}=1/3$ from
rotational invariance.}
\label{FIG13}
\end{figure}

We note that while the picture of the spins ``lying'' down in
the $xy$ plane is a physically appealing one, it is not in
fact so accurate a view quantitatively.  Fig.~13 shows the
square of the spin components as a function of field strength.
We show only one lattice size, 32x32, since there is no dependence on lattice
size for such a local quantity.
It is apparent that even for $h_{z}=2.0$ where Fig.~12
shows a very significant enhancement of the $xy$ plane
structure factor, the spins locally still point almost as much in the
$z$ direction as for $h_{z}=0.0$.  Indeed for larger fields,
the spins start to align ferromagnetically 
with the field and $\langle s_{z}^{2} \rangle$
exceeds $\langle s_{x}^{2} \rangle=\langle s_{y}^{2} \rangle$.
How do we reconcile this with the results for the structure factor?
Apparently, the uniform field $h_{z}$ has a large effect on the long range
spin correlations, and very little effect on the short range
correlations.  Thus while individual spins still rotate significantly in
the $z$ direction, the antiferromagnetic correlations in the $z$
direction between different spins are destroyed.  (In fact they
become ferromagnetic.)  At the same time, long range antiferromagnetic
correlations in the plane are enhanced, even though the individual spins are
not really ``lying down'' in the plane.

\begin{figure}
\begin{center}
\leavevmode
\hbox{%
\epsfxsize=3.25in
\epsffile{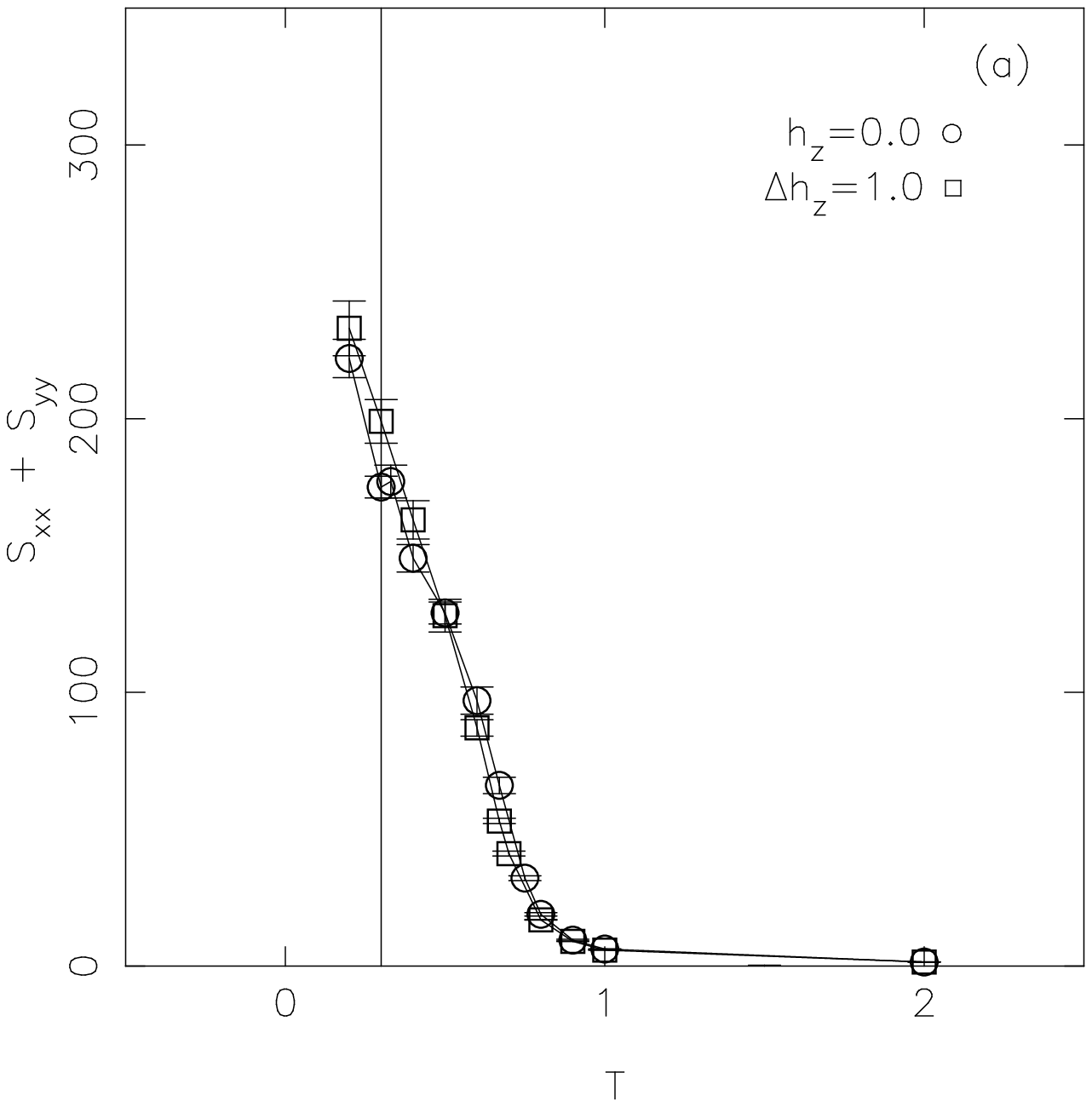}}
\end{center}
\end{figure}
\begin{figure}
\begin{center}
\leavevmode
\hbox{%
\epsfxsize=3.25in
\epsffile{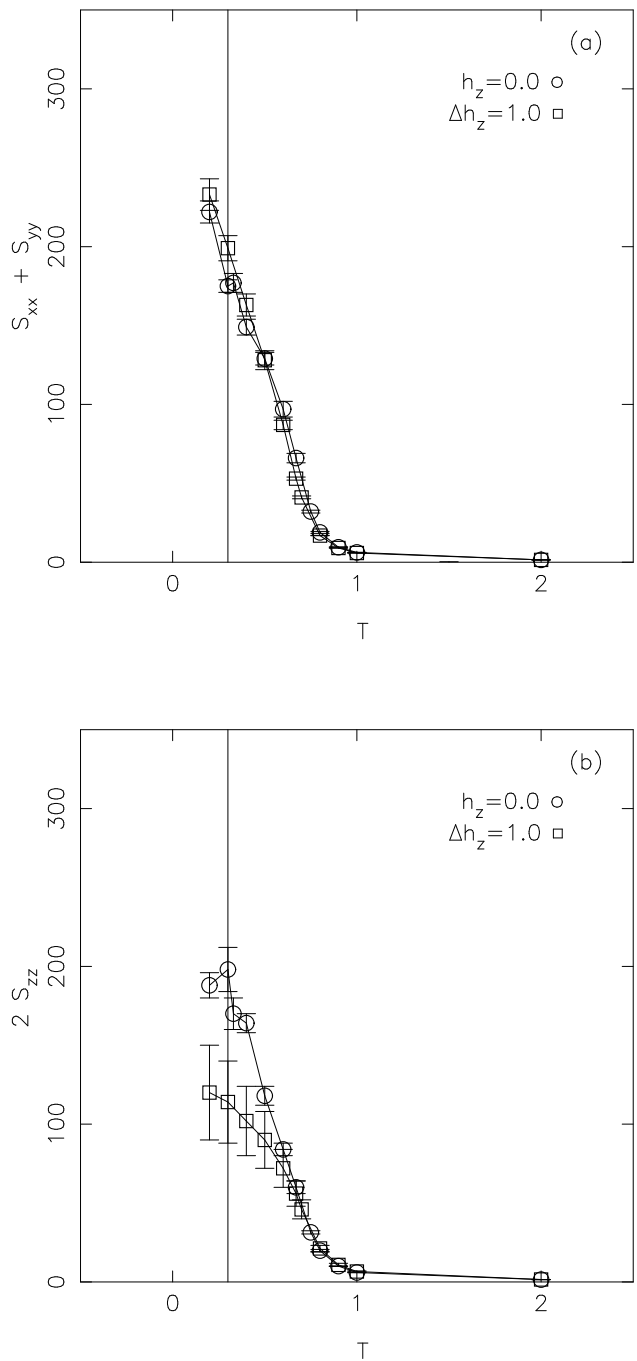}}
\end{center}
\caption{Results for the structure factors as a function of temperature
for a 20x20 lattice in both a zero field ($h_z = 0.0 $) and random field
($\Delta h_z = 1.0$).  (a) $S_{xx} + S_{yy};$ (b) $2 S_{zz}.$  The line at T=0.3 is at the temperature used in the finite size scaling plots (Fig.~15) below.}
\label{FIG14}
\end{figure}

\begin{figure}
\begin{center}
\leavevmode
\hbox{%
\epsfxsize=3.25in
\epsffile{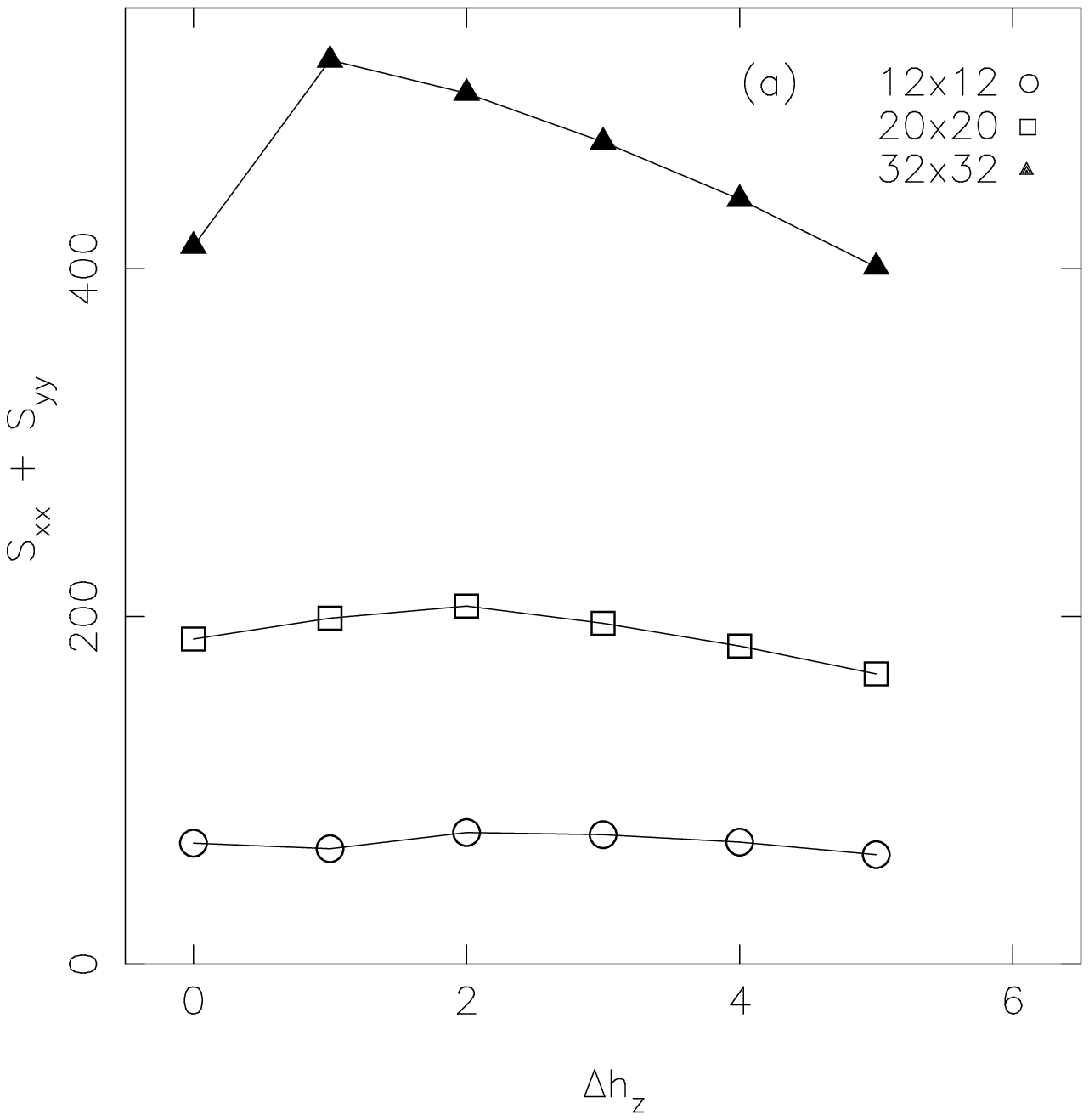}}
\end{center}
\end{figure}
\begin{figure}
\begin{center}
\leavevmode
\hbox{%
\epsfxsize=3.25in
\epsffile{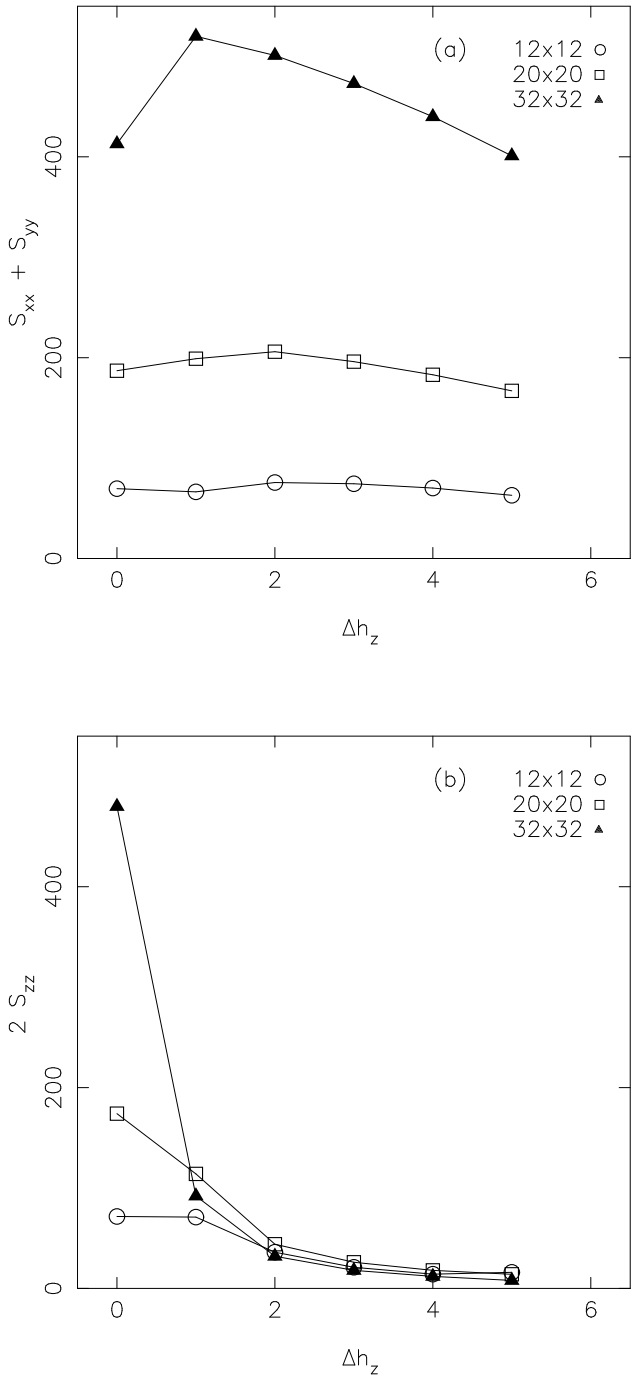}}
\end{center}
\caption{Results for the structure factors at T=0.3 as a function of random
field strength $\Delta h_{z}$ for different lattice sizes.  
(a) $S_{xx} + S_{yy};$ (b) $2 S_{zz}.$}
\label{FIG15}
\end{figure}

We now turn to the results of simulations for {\it random} magnetic
fields $h_{z}({\bf j})$, which is the situation to which
our random site energy attractive Hubbard model corresponds.
Figs.~14-15 show the analogous plots to those of the uniform field case.
Again for low $T$ a random field enhances $S_{xx}$ and $S_{yy}$,
although the effect is much less dramatic than seen 
in Fig.~11a.  The suppression
of $S_{zz}$ is more substantial, but again less decisive than with a uniform
field, Fig.~11b.
The random-field results shown in Fig.~15 were obtained
at T=0.3 while the results in the corresponding uniform-field Fig.~12 were
obtained at T=0.67, the
temperature at which previously reported uniform-field Heisenberg work was
done.\cite{SCALETTAR1}  For intermediate temperatures like $T = 0.67$
it was found that
$S_{zz}$ {\it exceeds,} by a small amount,
$S_{xx}$ and $S_{yy}$ for small disorder.  It is not clear to us why
this occurs. As seen by comparing Figs.~11b and 14b, $S_{zz}$ behaves
quite differently when the system is subjected to a uniform as opposed to
a random field along the $z$ axis.  A uniform field suppresses
$S_{zz}$ much more strongly than a random field.  Nevertheless,
in both cases the $z$ axis structure factor is suppressed with respect to the
$xy$ plane structure factors at low enough temperatures.

In Fig.~15, at T=0.3, we see that increasing the disorder suppresses the structure factor in the $z$ direction, the analogue of the cdw structure factor (Fig.~15b), while the structure factors in the $xy$ plane, the analogue of the pair
structure factor, remain robust to randomness (Fig.~15a).  Hence, we see the
same general results as seen in the disordered, attractive Hubbard model.  The analogy is somewhat limited,
since we do not see the destruction of the $xy$ plane structure
factor (Fig.~15a) as disorder is increased,
the analogue of the destruction of superconductivity (Fig.~6).
Determining the critical properties of the random field Heisenberg model
is a problem far beyond the scope of this paper.  However,
we note that a finite $V_c$ for $S_{xx}$ and $S_{yy}$
is not forbidden, for example, by an Imry-Ma argument,
since although the order parameter in this case has continuous symmetry, it
is not conjugate to the random disordered $h_z$ field.\cite{IMRY1}  

\section{Conclusions}

We have studied the effect of random site
energies on charge density wave and superconducting correlations in
the 2D attractive Hubbard Hamiltonian at half-filling.
We find an immediate suppression of cdw order with a random
one-body potential $V$, whereas the pair order is relatively robust.
This is consistent with a strong coupling picture of the
real space configurations of low energy states of 
pairs as disorder is turned on.

After showing the effect of $V$ on the spatial correlations,
we performed a finite size scaling analysis of the structure factors
and determined that long range cdw is destroyed for $V_{c} \approx 0.0$
while superconducting order has $V_{c} \approx 1.5$ which is roughly the
energy scale $4 t^2 / U \approx 1$ which stabilizes pairing.  It is
interesting that $V_c$ is considerably less than that required for the
destruction of superconductivity in the doped system.\cite{TRIVEDI3}

We discussed briefly the distribution of different measurements
as the disorder configuration was modified.  This enabled
us to determine a reasonable number of disorder realizations to average over.
Finally, the strong coupling picture of the classical
Heisenberg Hamiltonian in a random
external field $h_{z}({\bf j})$ was shown to 
exhibit some similarities with
the disordered attractive Hubbard model results at $|U|=4.0$.

There are a number of further questions we wish to explore.  First, 
we always performed our simulations at an inverse temperature $\beta$ 
sufficiently large so that we were in the ground state of the
finite size lattices we were studying.  The temperature
dependence of various quantities would be interesting to determine.
We would also like to look more at the non-equal time, dynamical
response of the system.  In particular, the issue of how
the superconducting gap fills in with disorder is an
interesting one.  This work requires analytic continuation of the
imaginary time Green's function, a task which should be challenging,
especially in the presence of disorder.

\acknowledgements
We gratefully acknowledge the help of Prof.~Mohit Randeria, 
Dr.~Karl~Runge, who was supported
by the National Science Foundation, grant ASC-9405041, and
Prof.~Nandini Trivedi,
and the support of National Science Foundation,
grant NSF-DMR-9520776 (RTS),
and Office of Naval Research, 
grant ONR N00014-93-1-0495 (CH), as well as time at the San Diego
Supercomputing Center.


\begin{references}
\bibitem{ERRORBARS}
Though our error bars do not make possible a definitive statement
on this point, it is
interesting that there appears a small initial enhancement of the order
parameter with $V$.  Similar effects have been seen in the behavior
of the staggered magnetization in coupled AF Heisenberg layers, where the
order parameter first slightly increases before interplanar coupling
drives a paramagnetic phase.\cite{SANDVIK}  It is also known that $T_N$ can
be enhanced by site impurities in the repulsive Hubbard model.\cite{ULMKE}
\bibitem{CAVEAT}
These numbers differ from those reported in \cite{SCALETTAR1},
for $h_z \approx 3 - 4,$ 
which were obtained with a Langevin rather than a Monte Carlo simulation.
We believe the present values to be correct.
In any case, the qualitative results are the same.
\end{references}
\end{document}